\documentclass[preprint,aps,prc,tightenlines,floatfix,showpacs,showkeys,nofootinbib]{revtex4}

\usepackage{graphicx}
\usepackage{dcolumn}
\usepackage{bm}
\usepackage{longtable}

\newcommand{\be}{\begin{equation}}
\newcommand{\ee}{\end{equation}}
\newcommand{\bea}{\begin{eqnarray}}
\newcommand{\eea}{\end{eqnarray}}

\begin{document}
\title{Neutrinoless double beta decay of $^{150}$Nd 
accounting for deformation}

\author{Dong-Liang Fang}
\affiliation{Institut f\"ur Theoretische Physik, Universit\"at T\"ubingen, 
D-72076 T\"ubingen, Germany}
\author{Amand Faessler}
\affiliation{Institut f\"ur Theoretische Physik, Universit\"at T\"ubingen, 
D-72076 T\"ubingen, Germany}
\author{Vadim Rodin}
\email{vadim.rodin@uni-tuebingen.de}
\affiliation{Institut f\"ur Theoretische Physik, Universit\"at T\"ubingen, 
D-72076 T\"ubingen, Germany}
\author{Fedor \v Simkovic}
\affiliation{BLTP, JINR, Dubna, Russia and Department of Nuclear Physics, Comenius University, SK-842 15 Bratislava, Slovakia}

\date{\today}

\begin{abstract}
A microscopic state-of-the-art calculation of the nuclear matrix element for neutrinoless double beta decay of $^{150}$Nd with an account for nuclear deformation is performed.
The proton-neutron quasiparticle random phase approximation (QRPA) with a realistic residual interaction [the Brueckner $G$ matrix derived from the charge-depending Bonn (Bonn-CD) nucleon-nucleon potential] is used as the underlying nuclear structure model. 
The present calculated matrix element is suppressed by about 40\% as compared with our 
previous QRPA result for $^{150}$Nd obtained with neglect of deformation.
By making use of this newest nuclear matrix element, one may conclude that 
neutrinoless double beta decay of $^{150}$Nd, to be measured soon by the SNO+ collaboration, provides one of the best probes of the Majorana neutrino mass.
\end{abstract}



\date{\today}
\maketitle

Neutrinoless double beta decay ($0\nu\beta\beta$ decay) is a
second order nuclear weak decay process with the emission of two electrons
only~\cite{Vog92,fae98,AEE07}: $(A,Z)\rightarrow
(A,Z+2)+2e^-$. This process violates the total lepton number conservation and is
therefore forbidden in the standard model (SM) of electroweak interaction. 
The existence of $0\nu\beta\beta$ decay will immediately prove the neutrino to be a Majorana particle (i.e., identical to its antiparticle). 

Furthermore, a study of $0\nu\beta\beta$ decay is an indispensable mean to probe the absolute neutrino masses at the level of tens of meV. 
The fact that the neutrinos are massive particles was firmly established by neutrino oscillation experiments, 
thus providing the first evidence for physics beyond the SM
(for reviews see, e.g., Ref.~\cite{Kay08}).
However, the observed oscillations cannot in principle pin down the absolute scale of the neutrino masses. This calls for alternative ways one of which is $0\nu\beta\beta$ decay.

Thus unambiguous observation of $0\nu\beta\beta$ decay would be of paramount importance for our understanding of particle physics beyond the SM.
The next generation of $0\nu\beta\beta$-decay experiments (CUORE, GERDA, MAJORANA, SNO+, SuperNEMO, and so on, 
see, e.g., Ref.~\cite{AEE07}  for a recent review) has a great discovery potential. 
Provided the corresponding decay rates are accurately measured, 
knowledge of the relevant nuclear matrix elements (NME) $M^{0\nu}$ will become indispensable to reliably deduce the effective Majorana mass 
from half-lives $T^{0\nu}_{1/2}$ of the decay.

One of the best candidate nuclei for searching $0\nu\beta\beta$ decay is $^{150}$Nd since 
it has the second highest endpoint, $Q_{\beta\beta}=$3.37 MeV, and the largest phase-space factor for the decay (about 33 times larger than that for $^{76}$Ge, see, e.g.~\cite{Vog92}). 
The SNO+ experiment at the Sudbury Neutrino Observatory 
will use an Nd-loaded scintillator to search for $0\nu\beta\beta$ decay by looking for a distortion in the energy spectrum of decays at the endpoint~\cite{SNO+}. 
SNO+ will be filled with 780 tons 
of liquid scintillator. The planned loading of 0.1\%  of the natural Nd translates into 43.6 kg of the isotope $^{150}$Nd. It is expected to achieve the sensitivity of $T^{0\nu}_{1/2} \simeq 5\cdot 10^{24}$ years after one year of running, with the best final value of about three to four times longer (without enrichment of the dissolved Nd). 

Now, to translate the anticipated experimental sensitivity to the decay rate into 
the sensitivity expected for the effective Majorana neutrino mass $m_{\beta\beta}$, 
one needs the corresponding NME $M^{0\nu}$. With the result $M^{0\nu}=4.74$ of Ref.~\cite{Rod05}, already the initial phase of SNO+ will be able to probe $m_{\beta\beta}\approx$ 100 meV, and will finally be able to achieve sensitivity of $m_{\beta\beta}\approx$ 50 meV corresponding to the case of the so-called inverse hierarchy (IH) of the neutrino mass spectrum.

However, $^{150}$Nd is well-known to be strongly deformed, which 
strongly hinders a reliable theoretical evaluation 
of the corresponding $0\nu\beta\beta$-decay NME.
For instance, it does not seem feasible in the near future to reliably treat this nucleus within the large-scale nuclear shell model (LSSM), see, e.g., Ref.~\cite{men09}. Also, the ``optimistic'' NME of Ref.~\cite{Rod05} was obtained within a microscopic approach, the proton-neutron quasiparticle random phase approximation (QRPA), with neglect of deformation.

Recently, more phenomenological approaches like the pseudo-SU(3) model~\cite{Hir95}, the Projected Hartree-Fock-Bogoliubov (PHFB) approach~\cite{Hir08} and the interacting boson model (IBM-2)~\cite{Bar09} have been employed to calculate 
$M^{0\nu}$ for strongly deformed heavy nuclei (a comparative analysis of different approximations involved in the models can be found in Ref.~\cite{Esc10}).
The results of these models 
generally reveal a substantial suppression of $M^{0\nu}$ for $^{150}$Nd as compared with the QRPA result of  Ref.~\cite{Rod05} where $^{150}$Nd and $^{150}$Sm were treated as spherical nuclei. 
The recent result of the PHFB~\cite{Hir08} is in a fair agreement with the  pseudo-SU(3) one of Ref.~\cite{Hir95}, but they both are about 1.5 times smaller than  $M^{0\nu}$ of the IBM-2~\cite{Bar09}. These results for $M^{0\nu}$ give a factor of 2--3 worse limits (as compared with the result of Ref.~\cite{Rod05}) on the Majorana neutrino mass to be achieved at SNO+ , and basically leave no hope to probe the IH region by the current configuration of SNO+. 

Such a spread in calculated NME $M^{0\nu}$ for $^{150}$Nd makes it very important to have a reliable estimate of the effect of nuclear deformation on $M^{0\nu}$. The most microscopic way up-to-date to describe this effect in $^{150}$Nd and $^{150}$Sm is provided by the QRPA. In Refs.~\cite{Sim03,Sim04,Sal09} a QRPA approach for calculating the $2\nu\beta\beta$-decay NME $M^{2\nu}$ in deformed nuclei was developed. The $2\nu\beta\beta$-decay half-lives have accurately been measured for a dozen nuclei and the corresponding nuclear matrix elements
$M^{2\nu}_{exp}$ were extracted~\cite{Bar10}. A theoretical interpretation of
these matrix elements provides a check of the reliability of different models. 
It was demonstrated in Refs.~\cite{Sim03,Sim04,Sal09} that 
deformation introduces a mechanism of suppression of the $M^{2\nu}$ matrix element 
which gets stronger when deformations of the initial and final nuclei differ from each other. A similar dependence of the suppression of both $M^{2\nu}$ and $M^{0\nu}$ matrix elements on the difference in deformations was found in the PHFB~\cite{Hir08} and the LSSM~\cite{men09}. 

In this Rapid Communication we report on the most 
microscopic state-of-the-art calculation of $M^{0\nu}$ for $^{150}$Nd with an account for nuclear deformation. The QRPA with a realistic residual interaction (the Brueckner $G$-matrix derived from the Bonn-CD nucleon-nucleon potential)~\cite{Sal09} is used.
The present calculation shows a suppression of $M^{0\nu}$ by 
about 40\% as compared with our previous QRPA result~\cite{Rod05} 
for $^{150}$Nd that was obtained with neglect of deformation.
Making use of this newest NME, one may conclude that $0\nu\beta\beta$ decay
of $^{150}$Nd, to be searched for by the SNO+ collaboration soon, provides one of the best sensitivities to the Majorana neutrino mass and may approach the IH region of the neutrino mass spectrum.

The NME $M^{0\nu}$ for strongly deformed, axially symmetric nuclei can be most conveniently calculated within the QRPA in the intrinsic coordinate system associated with the rotating nucleus. This employs the adiabatic Bohr-Mottelson approximation that is well justified for $^{150}$Nd, which indeed reveals strong deformation. As for $^{150}$Sm, the enhanced quadrupole moment of this nucleus is an indication for its static deformation. Nevertheless, the experimental level scheme of $^{150}$Sm does not reveal a clear ground-state rotational band. In this work we treat $^{150}$Sm in the same manner as $^{150}$Nd. 
However, a more elaborated theoretical treatment going beyond the simple adiabatic approximation might be needed in the future to describe the nuclear dynamics of this nucleus.

Nuclear excitations in the intrinsic system $| K^\pi\rangle$ are characterized by the projection $K$ of the total angular momentum onto the nuclear symmetry axis (the only projection that is conserved in strongly deformed nuclei) and the parity $\pi$. 
In Ref.~\cite{Sal09} the structure of the intermediate 
$| 0^+\rangle$ and $| 1^+\rangle$ states was obtained within the QRPA to calculate $2\nu\beta\beta$-decay NME $M^{2\nu}$. Here, the approach of Ref.~\cite{Sal09} is straightforwardly extended to calculate all possible $| K^\pi\rangle$ states needed to construct $M^{0\nu}$.

The matrix element $M^{0\nu}$ is given within the QRPA in the intrinsic system by a sum of the partial amplitudes of transitions via the intermediate states $K^\pi$
\begin{equation}
M^{0\nu}=\sum_{K^\pi} M^{0\nu}(K^\pi)\ ; \ M^{0\nu}(K^\pi) = 
\sum_{\alpha} s^{(def)}_\alpha O_\alpha(K^\pi). \label{M0n}
\end{equation}
Here we use the notation of Appendix B in Ref.~\cite{anatomy}, $\alpha$ stands for the set of four single-particle indices $\{p,p',n,n'\}$, 
and $O_\alpha(K^\pi)$ is a two-nucleon transition amplitude via all the $K^\pi$ states
 in the intrinsic frame
\begin{equation}
O_\alpha(K^\pi)=\sum_{m_i,m_f}
\langle 0_f^+|c_{p}^\dagger c_{n}|K^\pi m_f\rangle 
\langle K^\pi m_f|K^\pi m_i\rangle
\langle K^\pi m_i|c^\dagger_{p'} c_{n'}|0_i^+\rangle .
\label{O}
\end{equation}
The two sets of intermediate nuclear states generated from the
initial and final ground states (labeled by $m_i$ and $m_f$, respectively) 
do not come out identically within the
QRPA. A standard way to tackle this problem is to introduce the overlap factor of these states $\langle K^\pi m_f|K^\pi m_i\rangle$ in Eq.~(\ref{O}).
Two-body matrix elements $s^{(def)}_\alpha$ of the neutrino potential in 
Eq.~(\ref{M0n}) in a deformed Woods-Saxon single-particle basis are decomposed over the the spherical harmonic oscillator ones according to the way described in Ref.~\cite{Sal09}:
\begin{equation}
s^{(def)}_{pp'nn'}=
\sum_{J}\sum_{\footnotesize\begin{array}{c}
\eta_p \eta_{p'}\\[-1pt]  \eta_n \eta_{n'}\end{array}}
F^{JK}_{p\eta_p n\eta_n}F^{JK}_{p'\eta_{p'}n'\eta_{n'}}s^{(sph)}_{\eta_p\eta_{p'} \eta_n\eta_{n'}}(J),
\end{equation}
\begin{eqnarray}
s^{(sph)}_{pp'nn'}(J)&=&\displaystyle \sum_{\mathcal J}
(-1)^{j_n + j_{p'} + J + {\mathcal J}} \hat{\mathcal J}
\left\{
\begin{array}{c c c}
j_p & j_n & J \\ j_{n'} & j_{p'} & {\mathcal J}
\end{array}
\right\} 
\langle p(1), p'(2); {\mathcal J} \| {\mathcal O_\ell}(1,2) \| n(1), n'(2); {\mathcal J} \rangle\,,
\end{eqnarray}
where $\hat{\mathcal{J}} \equiv \sqrt{2\mathcal{J}+1}$, and ${\mathcal O_\ell}(1,2)$ is the neutrino potential as a function of the coordinates of two particles, with ${\ell}$ labeling its Fermi (F), Gamow-Teller (GT), and Tensor (T) parts. The particle-hole transformation coefficient 
$F^{JK}_{p\eta_p n\eta_n}= B^p_{\eta_p}B^n_{\eta_n}(-1)^{j_n-\Omega_{n}}C^{JK}_{j_p\Omega_{p} j_n-\Omega_{n}}$ from the deformed basis into the spherical harmonic oscillator one
is constructed from the single-particle decomposition coefficients $B^p_{\eta_p}$ and $B^n_{\eta_n}$ (see Ref.~\cite{Sal09} for details), 
$C^{JK}_{j_p\Omega_{p} j_n-\Omega_{n}}$ is the Clebsch-Gordan coefficient.

The particle-hole transition amplitudes in Eq.~(\ref{O}) can be represented in terms 
of the QRPA forward $X^m_{i K}$ and backward $Y^m_{i K}$  amplitudes along with 
the coefficients of the Bogoliubov transformation $u_\tau$ and $v_\tau$~\cite{Sal09}:
\begin{eqnarray}
\langle 0_f^+|c_{p}^\dagger c_{n}|K^\pi m_f\rangle&=&v_{p}u_{n}X^{m_f}_{pn,K^\pi}+u_{p}v_{n}Y^{m_f}_{pn,K^\pi},\nonumber\\
\langle K^\pi m_i|c^\dagger_p c_{n}|0_i^+\rangle&=&u_{p}v_{n}X^{m_i}_{pn,K^\pi}+v_{p}u_{n}Y^{m_i}_{pn,K^\pi}.\nonumber
\end{eqnarray}
The overlap factor in Eq.~(\ref{O}) can be written as:
\begin{eqnarray}
\langle K^\pi m_f|K^\pi m_i\rangle&=&\sum_{l_il_f}[X^{m_f}_{l_fK^\pi}X^{m_i}_{l_iK^\pi}-Y^{m_f}_{l_fK^\pi}Y^{m_i}_{l_iK^\pi}]
\mathcal{R}_{l_fl_i}\langle BCS_f|BCS_i\rangle
\label{overlap}
\end{eqnarray}
Representations for ${\cal R}_{l_fl_i}$  and the overlap factor $\langle BCS_f|BCS_i\rangle$ between the initial and final BCS vacua  are given in Ref.~\cite{Sim03}.


For a numerical computation of the $0\nu\beta\beta$-decay
NME  $M^{0\nu}$ for the process $^{150}$Nd$\rightarrow ^{150}$Sm$+2e^-$, we have 
straightforwardly extended the method of Ref.~\cite{Sal09}.

The single-particle Schr\"odinger equation with the Hamiltonian of 
a deformed Woods-Saxon mean field is solved on the basis of a axially-deformed 
harmonic oscillator. The parametrization of the mean field is adopted from the spherical
calculations of Refs.~\cite{Rod05,anatomy,Sim09}.  
We use here the single-particle deformed basis corresponding in the spherical limit to full (4--6)$\hbar\omega$ shells.
Decomposition of the deformed single-particle wave functions is performed over the spherical harmonic oscillator states within the seven major shells.
Only quadrupole deformation is taken into account in the calculation. The geometrical quadrupole deformation parameter $\beta_2$ of the deformed Woods-Saxon mean 
field is obtained 
by fitting the experimental deformation parameter $\beta= \sqrt{\frac{\pi}{5}}\frac{Q_p}{Z r^{2}_{c}}$, where $r_c $ is the charge rms radius and  $Q_p$ is the empirical intrinsic quadrupole moment. The latter  
can be derived from the laboratory quadrupole moments measured by the Coulomb excitation reorientation technique, or from the corresponding 
$B(E2)$ values~\cite{ragha}. 
We take in this work the experimental values $\beta=0.29$ and 
$\beta=0.19$ for $^{150}$Nd and $^{150}$Sm, respectively, which are 
extracted from the $B(E2)$ values as being more accurate.
The fitted values of the parameter $\beta_2$ of the deformed Woods-Saxon mean 
field, which allow us to reproduce the experimental $\beta$, are listed in Table~\ref{table.1}. The spherical limit, i.e. $\beta_2=0$, is considered as well, to compare with the earlier results of Ref.~\cite{Rod05}. The  procedure of fitting $\beta_2$ adopted here is more consistent than the approximate ansatz $\beta_2=\beta$ used in Ref.~\cite{Sal09}.

As in Refs.~\cite{Rod05,Sal09,anatomy,Sim09}, the nuclear Brueckner $G$ matrix, obtained by a solution of the Bethe-Goldstone equation with the Bonn-CD one boson exchange nucleon-nucleon potential, is used as a residual two-body interaction. 
First, the BCS equations are solved to obtain the Bogoliubov coefficients, gap parameter and the chemical potentials. To solve the QRPA equations,
one has to fix the particle-hole $g_{ph}$ and particle-particle
$g_{pp}$ renormalization factors of the residual interaction (see Ref.~\cite{Sal09} for details). As in  Ref.~\cite{Sal09}, we determine a value of $g_{ph}$ 
by fitting the experimental  position of the Gamow-Teller giant resonance (GTR) in the intermediate nucleus. Since there is no experimental information on the GTR energy for $^{150}$Nd, we use for this nucleus the same $g_{ph}=0.90$ as fitted for $^{76}$Ge (this value is slightly different from the fitted $g_{ph}=1.15$ of Ref.~\cite{Sal09} 
because of a different parametrization of the mean field used here). 
The parameter $g_{pp}$ can be determined by fitting the experimental value of the $2\nu\beta\beta$-decay NME $M^{2\nu}_{GT}=0.07$ MeV$^{-1}$~\cite{Bar10}. The unquenched axial-vector coupling constant $g_A=1.25$ is used here. The fitted values of $g_{pp}$ are listed Table~\ref{table.1}.
Note, that the more realistic procedure of fitting $\beta_2$ adopted here also gives us more realistic $g_{pp}\simeq 1$ values as compared with those of Ref.~\cite{Sal09}.

\begin{table}[h]
\centering
\caption{The values of the deformation parameter $\beta_2$ of Woods-Saxon mean field 
 for initial (final) nuclei fitted in the calculation to reproduce the experimental quadrupole moment. 
Also the fitted values of the particle-particle strength parameter $g_{pp}$ are listed 
(the particle-hole strength parameter is $g_{ph}=0.90$).
The BCS overlap factor $\langle BCS_f|BCS_i\rangle$ (\ref{overlap}) between the initial and final BCS vacua is given in the last column.}
\label{table.1}
\begin{tabular}{|l|c|c|c|c|}
	\hline
 & & &  \\
initial (final)&  $\beta_{2}$&\ $g_{pp}$\ 
&$\langle BCS_i|BSC_f\rangle$\\	
nucleus & & &  \\	

\hline
 & & &  \\ 

$^{150}$Nd ($^{150}$Sm)& 0.240 (0.153) & 1.05 
& 0.52 \\
                       &  0.0\ \  (0.0) & 1.01 
& 0.85 \\ 
\hline 
\end{tabular}
\end{table}

Having solved the QRPA equations, the two-nucleon transition amplitudes (\ref{O}) are calculated and, by combining them with the two-body matrix elements of the neutrino potential, the total $0\nu\beta\beta$ NME $M^{0\nu}$ (\ref{M0n}) is formed. The present computation is rather time consuming since numerous programming loops are needed to calculate the decompositions of the deformed two-body matrix elements over the spherical ones. Therefore, to speed up the calculations the mean energy of 7 MeV of the intermediate states is used in the neutrino propagator.
Following Refs.~\cite{Rod05,anatomy,Sim09}, in this first application of the approach
the effects of the finite nucleon size and higher-order weak currents are included.
Recently, it was shown~\cite{Sim09} that a modern self-consistent treatment of the two-nucleon short-range correlations (s.r.c.) leads to a change in the NME $M^{0\nu}$  only by a few percents, much less than the traditional Jastrow-type representation of the s.r.c. does. Therefore, we postpone the analysis of the anticipated small effects of the s.r.c. to a forthcoming detailed publication.

In Table~\ref{tab:3} the presently calculated NME $M^{0\nu}$ for $^{150}$Nd is listed (column 4) and compared with the calculation results by other approaches. One can see that the NME $M^{0\nu}$ of this work calculated with neglect of deformation (column 3) agrees well with the previous one of the spherical QRPA~\cite{Rod05}. A small difference can have its origin in the somewhat different approximations involved (use of the Woods-Saxon single particle wave functions and the BCS overlap factor, neglect of the s.r.c. in the present work). By including deformation (column 4), one gets about 1.8 times smaller NME $M^{0\nu}$. The main origin of the suppression can be attributed to a smaller BCS overlap factor in the latter case, that is due to a marked difference in deformations between $^{150}$Nd and $^{150}$Sm nuclei (see Table~\ref{table.1}).

Our present NME $M^{0\nu}$ for $^{150}$Nd, obtained within the state-of-the-art QRPA approach that accounts for nuclear deformation, though smaller than the earlier one of Ref.~\cite{Rod05}, still is significantly larger than the NME of other approaches (columns 5,6, and 7 of Table~\ref{tab:3}). 
The $0\nu\beta\beta$-decay half-life $T^{0\nu}_{1/2}$ corresponding to the Majorana neutrino mass $\langle m_{\beta\beta} \rangle$ = 50 meV is more than two times shorter as compared with the most optimistic prediction of the IBM-2 among the other 
approaches~\footnote{Note that by neglecting the Jastrow-type s.r.c. the IBM-2 NME will get about 20\% larger and be in rather good agreement with our present result.}. 
It allows us to hope that the SNO+ experiment will still be able to approach the inverse hierarchy of the neutrino mass spectrum.

\begin{table*}
\caption{
The matrix elements $M^{0\nu}$ for the $0\nu\beta\beta$ decay $^{150}$Nd$\rightarrow ^{150}$Sm calculated in different models. The final result of this work obtained with account of deformation is given in column 4. A result with neglect of deformation is also listed (column 3) for comparison with the earlier result of Ref.~\cite{Rod05} (column 2). The corresponding half-lives $T^{0\nu}_{1/2}$ (in years) for an assumed effective Majorana neutrino mass $\langle m_{\beta\beta} \rangle$ = 50 meV are also shown.
%
}
\begin{tabular}{|ccccccc|} 
\hline
{} 
&  {QRPA~\cite{Rod05}
\footnote{using spherical harmonic oscillator
    wave functions, no 
deformation allowed. The radius parameter $r_0=1.2$ fm is used here, instead of $r_0=1.1$ fm of Ref.~\cite{Rod05}}}
&  {this work ($\beta_2=0$) 
\footnote{using Woods-Saxon wave functions, no 
deformation allowed.}}
&  {\bf this work
}
&  {pseudo-SU(3)~\cite{Hir95}}
&  {PHFB~\cite{Hir08}}
&  {IBM-2~\cite{Bar09}}
\\
\hline  
$M^{0\nu}$   &  5.17 & 5.78 & {\bf 3.16} & 1.57 & 1.61 & 2.32 \\
$T^{0\nu}_{1/2}$, $10^{25}$ y & 1.72 & 1.38 & {\bf 4.60} & 18.7 & 17.7 & 8.54 \\
($\langle m_{\beta\beta} \rangle$ = 50 meV) &   &  &  &  &  &  \\
\hline
\end{tabular} 
\label{tab:3}
\end{table*}


To conclude, in this Rapid Communication the most  
microscopic state-of-the-art calculation of the nuclear matrix element for neutrinoless double beta decay of $^{150}$Nd with an account for nuclear deformation is performed.
The proton-neutron 
QRPA with a realistic residual interaction (the Brueckner $G$ matrix derived from the Bonn-CD nucleon-nucleon potential) is used as the underlying nuclear structure model. 
The $0\nu\beta\beta$ decay matrix elements $M^{0\nu}$ calculated in this work shows 
suppression by about 40\% with respect to our 
previous QRPA result for $^{150}$Nd obtained with neglect of deformation.
Making use of this newest nuclear matrix element, one may conclude that 
neutrinoless double beta decay of $^{150}$Nd, to be measured soon by the SNO+ collaboration, provides one of the best sensitivities for the Majorana neutrino mass.

The authors acknowledge the support of the Deutsche Forschungsgemeinschaft under both SFB TR27 "Neutrinos and Beyond" and Graduiertenkolleg GRK683.

\end{document}